\documentclass{emulateapj}
\usepackage{graphicx}
\usepackage{natbib}

\def\kms{\>{\rm km}\,{\rm s}^{-1}}

\def\Msun{\>{\rm M_{\odot}}}

\newcommand{\gtsim}{\mathrel{\hbox{\rlap{\lower.55ex \hbox {$\sim$}}
                   \kern-.3em \raise.4ex \hbox{$>$}}}}
\newcommand{\ltsim}{\mathrel{\hbox{\rlap{\lower.55ex \hbox {$\sim$}}
                   \kern-.3em \raise.4ex \hbox{$<$}}}}

\begin{document}

\title{Gravitational Wave Signal from Assembling the Lightest Supermassive Black Holes}

\author{Kelly Holley-Bockelmann\altaffilmark{1}, Miroslav
  Micic\altaffilmark{2}, Steinn Sigurdsson\altaffilmark{3}, and Louis
  J. Rubbo\altaffilmark{4}}
\altaffiltext{1}{Department of Physics and Astronomy, Vanderbilt University, Nashville, TN, 37235 Email: k.holley@vanderbilt.edu}
\altaffiltext{2}{Center for Gravitational Wave Physics,
The Pennsylvania State University, University Park, PA 16802
Email: steinn@astro.psu.edu}
\altaffiltext{3} {Department of Physics, University of Sydney , Sydney, Australia Email:micic@physics.usyd.edu.au}
\altaffiltext{4} {Department of Chemistry and Physics, Coastal Carolina University, Conway, SC 29528, lrubbo@coastal.edu}

\begin{abstract}
We calculate the gravitational wave signal from the growth of 
$10^7 \, {\rm M}_\odot$ supermassive black holes (SMBH) 
from the remnants of Population III stars. The assembly of
these lower mass black holes is particularly important because
observing SMBHs in this mass range is one of the primary science
goals for the Laser Interferometer Space Antenna (LISA), a planned NASA/ESA 
mission to detect gravitational waves. We use high resolution 
cosmological N-body simulations to track the merger history of the host
dark matter halos, and model the growth of the SMBHs with a semi-analytic
approach that combines dynamical friction, gas accretion, and feedback.
We find that the most common source in the LISA band from our volume
consists of mergers between intermediate mass black holes
 and SMBHs at redshifts less than 2.
This type of high mass ratio merger has not been widely considered in
the gravitational wave community; detection and characterization of
this signal will likely require a different technique than is used for
SMBH mergers or extreme mass ratio inspirals.  We find that the event
rate of this new LISA source depends on prescriptions for gas
accretion onto the black hole as well as an accurate model of the
dynamics on a galaxy scale; our best estimate yields $\sim$ 40 sources
with a signal-to-noise ratio greater than 30 occur within a volume
like the Local Group during SMBH assembly -- extrapolated over the
volume of the universe yields $\sim$ 500 observed events over 10
years, although the accuracy of this rate is affected by cosmic
variance.

\end{abstract}


\section{INTRODUCTION}

Any mass configuration that creates a time-dependent quadrupole of the
stress-energy tensor will radiate away energy in gravitational
waves. One of the best examples of this is the binary pulsar system
J1915+1606, whose measured orbital decay of $76 \, {\rm \mu s/yr} $ is
extremely well-described by gravitational wave
emission~\citep{hulse:75}.  However, even for very massive and compact
astrophysical objects, the strain amplitude of this fluctuation in
spacetime, $h$, is much less than $10^{-20}$. For this reason,
gravitational radiation has yet to be directly detected by current
ground-based gravitational wave observatories such as the LIGO, VIRGO,
and TAMA.

The strongest expected gravitational wave sources occur from the
inspiral and merger of binary supermassive black holes.  At the
current epoch, nearly every galaxy is thought to host a supermassive
black hole (SMBH) with a mass of $10^6 - 10^{10} \,
{\rm M}_\odot$~\citep[e.g.][]{kormendy1995}. Observational evidence suggests
that the black holes are embedded in galactic nuclei even at high
redshift, and grow more massive as the galaxy grows~\citep[e.g.][]{Soltan:1982:smbh,david:1987:eag, Silk:1998:smbh,kauffmann:2000:ume, Monaco:2000:smbh,Schneider:02, Granato:2001:smbh,merloni:2004:ags}.  Since galaxies
are thought to assemble by merging, each galaxy merger is expected to
spawn a binary SMBH~\citep{tamara09:bbh,Mayer:2007:smbh,Escala:2006:gas,Hopkins:2005:smbh,Wyithe:2005:msigma,Komossa:2002tn,Menou:01,Adams:2001:smbh,Haehnelt:2002:msb}. In general, the inspiral and coalescence of low
mass SMBHs are expected to spawn the largest amplitude gravitational
wave signal~\citep[][e.g.]{Haehnelt:1998}.  The lightest SMBH binaries, with masses $10^5-10^7 \,$
{\rm M}$_\odot$, have coalescence frequencies between $\sim 10^{-4} - 1 $  Hz -- squarely within the frequency range of the
NASA/ESA Laser Interferometer Space Antenna (LISA) which has been
proposed to launch in the next decade. LISA observations will
complement ground-based gravitational wave observatories by broadening
the observable gravitational wave spectrum to include low
frequencies. In particular, gravitational wave detections made by LISA
will be able to directly map the assembly of these lightest SMBHs~\citep{Haehnelt:94,Menou:01, Enoki:04}.

In the current picture of SMBH assembly, the 
black hole begins life as a low mass ``seed'' black hole at high redshift. 
It's not clear, though,  when exactly these BH seeds emerge or
what mass they have at birth. SMBH seeds may 
have been spawned from the accretion of low angular momentum gas in 
a dark matter halo~\citep{koushiappas:2004:mbh,Bromm:04}, 
the coalescence of many seed black holes within a 
halo~\citep{begelman:1978:fds, islam:04}, or from an IMBH formed, 
perhaps, by runaway stellar collisions~\citep{Portegies:2004fm, 
miller:2004:ibh, vandermarel:2004:ibh}. However, the most likely
candidates for SMBH seeds are the remnants that form from the first 
generation of stars sitting deep within dark matter 
halos~\citep{Madau:01Pop3, Heger:03sn, Volonteri:03smbh, 
Islam:03, Wise:05snpop3} -- so called Population III stars. 
With masses $< 10^3 \Msun$, these relic seeds are predicted to lie near 
the centers of dark matter halos between z $\sim 12-20$~\citep{Bromm:99, 
Abel:2000, Abel:02firststar}. Structure 
formation dictates that dark matter halos form in the early universe
and hierarchically merge into larger bound objects, so naturally as 
dark matter halos merge, seed black holes sink to the center through dynamical
friction and eventually coalesce. Dark matter halo mergers become 
synonymous, then, with black hole mergers at these masses and redshifts.
This means that although the seed formation stops at z$\sim$12 as 
Population III supernovae rates drop to zero~\citep{Wise:05snpop3}, 
SMBH growth continues as dark matter halo mergers proceed to low redshifts.

Gas accretion is thought to play a critical role in 
fueling the early stages of black hole growth~\citep{david:1987:eag, 
kauffmann:2000:ume, merloni:2004:ags}, and this may explain 
the tightness of the M$_{\rm BH}-\sigma$ relation~\citep{Burkert:2001:smbh,
haehnelt:2000:cbh, Dimatteo:2005, Kaz:2005:msigma, 
Robertson:2006:msigma}. Since high redshift galaxies are thought to be 
especially gas-rich, each merger brings a fresh supply of gas to the center 
of the galaxy, and new fuel to the growing supermassive black hole~\citep{Mihos:1994:gasmerger, Dimatteo:2003:bhgrowth}. From a combination 
of gas accretion and binary black hole coalescence, it is thought that 
these Pop III-generated seeds may form the SMBHs we observe 
today~\citep{Soltan:1982:smbh, Schneider:02topheavy}.

During a galaxy merger, each black hole sinks to the center of the new galaxy 
potential due to dynamical friction and eventually becomes bound as 
a binary~\citep{Kaz:2005:msigma, Escala:2005:gas}. Dynamical 
friction then continues to shrink the orbit until the binary is hard 
(i.e, the separation between each black hole, a$_{\rm BBH}$, is such that the 
system tends to lose energy during stellar encounters)~\citep{heggie:07}.
Thereafter, further decay is
mediated by 3-body scattering with the ambient stellar background
until the binary becomes so close that the orbit can lose energy
via gravitational radiation. In studies of static, spherical
potentials, it may be difficult for stellar encounters alone to 
cause the binary to transition between the 3-body scattering phase
and the gravitational radiation regime~\citep{milos:2003:lem}.
 However, in gas-rich or
non-spherical systems, the binary rapidly hardens and coalesces into one
black hole, emitting copious gravitational radiation in the process~\citep{Mayer:2007:smbh, Kaz:2005:msigma, berczik:2006:emb, sigurdsson:03, KHB:2006fl}.

In our previous work, we calculated the cosmological merger rate for 
black holes between 200 - $3 \times 10^7 \Msun$ from redshift 
49-0~\citep[][(hereafter, M07)]{KHB:2007bhgrowth, KHB:2008bhgrowth}. Our approach
combined high-resolution, small-volume cosmological N-body simulations
with analytic prescriptions for the dynamics of merging black holes below our
resolution limit; this allowed us to explore different black hole growth 
mechanisms and seed formation scenarios while also accurately simulating the
rich and varied merger history of the host dark matter halos.

In this paper, we calculate the gravitational wave signal from the 
black hole mergers involved in assembling a supermassive black hole 
at the center of the Milky Way analogue our simulation volume. The volume is designed to provide one possible evolutionary path
for a region like our Local Group, and as such, it should 
contain supermassive black holes on the light end of the supermassive 
black hole mass spectrum -- the sweet spot in SMBH mass for LISA observations.
We include all the mergers that have
occurred from redshift 49 to the present epoch within a 1000 Mpc$^3$ volume 
of the Universe that represents a Local Group
type of environment.

We found that the gravitational wave sources 
revealed in this volume are from much higher mass ratio mergers
that the mergers predicted to be involved in assembling the most massive
SMBHs. Most of the LISA science and data analysis community
has been anticipating more equal mass mergers, and have developed extensive
gravitational wave templates and parameter extraction techniques based on 
the assumption that the black hole binaries are 
order unity mass ratio systems \citep[][e.g.]{Babak:2008}. While this may
be true for the most massive SMBHs, we find that the black holes in our 
volume experience mergers with mass ratios as high as ${\rm M}_2/{\rm M}_1= 10000:1$. These high
mass ratio mergers certainly generate a different gravitational wave signal;
they may even deserve a different source classification to separate it
from the classical equal mass merger or extreme mass ratio inspiral\footnote{Extreme mass ratio inspirals involve compact objects, like white dwarfs, neutron stars, or stellar mass black holes spiraling into a SMBH.}. 

 We use the rates
found in our volume to extrapolate this gravitational wave signal 
over the Hubble Volume. Naturally, this is highly sensitive to
cosmic variance, and ten runs selected from an initial larger volume are planned to mitigate cosmic variance in
our merger rate estimates. However, these small volume cosmological simulations 
are extremely nonlinear, and as such, are computationally expensive. 
We are publishing the preliminary results from our first, smaller  
scale, simulation in order to draw the attention of the community 
to the interesting possibility that the assembly of the lowest mass 
SMBHs may involve these high mass ratio mergers and
be a significant contributor to the LISA detectable event rate. 


We review the details of our simulation in section 2 and in section 3, 
we describe how to calculate the gravitational wave signal from 
two merging black holes. We discuss our results and 
implications in section 4.

\section{Tracking the merger history of SMBHs}

In ~\citet{KHB:2008bhgrowth}, we performed a high-resolution 'zoom-in'
cosmological N-body simulation of a comoving section of 
a $\Lambda$CDM universe ($\Omega_{\rm M}$=0.3, $\Omega_{\Lambda}$=0.7, 
$\sigma_8$=0.9 and h=0.7) from z$=49$ to  z$=0$. The high resolution region 
was a box 10 $h^{-1}$ Mpc on each side, for a total comoving volume of 
1000 $h^{-3}$ Mpc$^3$. The volume is designed to provide one possible evolutionary path
for a region like our Local Group, and as such, it should 
contain supermassive black holes on the light end of the supermassive 
black hole mass spectrum.

Our mass 
resolution is $8.85 \times 10^5 \, {\rm M}_\odot$, and our spatial 
resolution is 2 kpc. After the simulation is complete, we identified halos
with at least 32 particles using P-Groupfinder~\citep{pgroupfinder} 
and seed black holes those halos in the appropriate mass and redshift 
range to host Pop III stars.  Note that we are using 
WMAP3~\citep{spergel:07} cosmological parameters in this study to
compare with our previous work; however, at the time of this paper's 
submission, 'zoom-in' simulations of several small volumes are underway with 
WMAP5 parameters to better explore cosmic variance.
This will allow us to better pin down the rate of these new gravitational 
wave sources. 

In our hybrid method, we combine the dark matter halo merger trees obtained in 
numerical simulations with an analytical treatment of the physical processes 
that arise in the dynamics of galaxy and black hole mergers. Since 
some of the processes are ill-constrained, we probe the effect of different
black hole growth recipes on the final black hole mass function. In general, 
we assume 
each dark matter halo is described by an NFW profile~\citep{nfw:97}, and include the effects of dynamical friction and 
merger-induced gas accretion onto the SMBH, as well as the SMBH 
merger itself. This hybrid approach generates, for each recipe, a 
census of the number and mass ratio of the mergers in our volume at 
each redshift.

One of the surprises from this method is that the black hole mass function is
quite sensitive to the dynamics of the host dark matter halo on large scales.
Previous semi-analytic work on SMBH growth all assume that the host galaxies 
assembled from binary mergers whose timescale was dictated by 
Chandrasekhar dynamical friction~\citep{chandra:43}. Numerical simulations, however,
indicate that galaxies can often assemble from multiple near-simultaneous 
mergers, and that the merger timescale is longer than what the Chandrasekhar
dynamical friction timescale suggests. Simulations indicate that the
merger timescale is only smaller by a factor 3 for minor mergers at z=0~\citep{Boylan:08}, 
so given the myriad uncertainties in the dynamics acting in the other stages 
of the merger, this factor of a few was neglected. However, a longer 
dynamical friction timescale reduces the total number of black hole
mergers in a volume of the universe by pushing some mergers to the
future. For example, we found that the Boylan-Kolchin treatment for dynamical friction resulted
in 1056 SMBH mergers in our volume, while Chandrasekhar dynamical friction
yielded 1245 mergers. In addition, 
most of the current work on SMBH evolution indicates that gas accretion
dominates the black hole's growth and that this gas is funneled onto the SMBH
by the merger process itself~\citep[e.g.][]{Robertson:2006:msigma,Hopkins:2005:smbh,Dimatteo:2003:bhgrowth,Volonteri:2002am}; if the gas is driven to the center over a 
longer timescale, then each merger may excite a longer gas-fueled growth spurt. In the context of gravitational wave source
prediction, these two effects may combine to make the SMBH mergers {\it louder} and
at {\it lower frequency} than what has been predicted with a Chandrasekhar dynamical
friction prediction. We will return to this in Section 4.

In this paper, we select a sample of the black hole growth recipes from~\citet{KHB:2008bhgrowth} and calculate the gravitational
wave signal from each of the black hole mergers in the volume. Briefly, the recipes span two choices for the dynamical
friction timescale, and three choices 
for the mass ratio of the halos that excite merger-driven gas accretion 
onto the black hole (4:1, 10:1, and $\infty:1$ -- a.k.a. 'dry growth').

The next section outlines this black hole growth prescription.

\subsection{SMBH Growth Prescription}

The SMBH in our model grows through a combination of black hole 
mergers and gas accretion. To better separate the effects
of gas accretion on the black hole, we include a dry growth scenario, 
where the black hole grows through mergers only.   At high redshift, 
this galaxy merger-driven approach is likely a good assumption, though note that at low redshift when
mergers are infrequent, secular evolution, such as bar instabilities, 
may dominate the gas (and therefore black hole) accretion. Integrated over 
the whole of a black hole lifetime, though, this
major merger-driven accretion is likely to be the dominant source of 
gas inflow. Since the black hole growth is so strongly dependent on what 
fuel is driven to the center during galaxy mergers,
it is important to characterize this merger-driven gas inflow, including 
the critical gas physics that may inhibit or strengthen this nuclear supply.
We are motivated by a recent suite of numerical simulations that include 
radiative gas cooling, star formation, and stellar feedback to study the 
starburst efficiency for unequal mass ratio galaxy mergers~\citep{Cox:2008}, 
which finds that the gas inflow depends strongly on the mass ratio of the 
galaxy~\citep[see also, e.g.][]{hernquist:89, Mihos:1994:gasmerger}.
This study parametrizes the efficiency of nuclear star formation 
(i.e. gas supply and inflow), $\alpha$, as a function of galaxy mass ratio:

\begin{equation}
{\alpha} = \Bigg({M_{\rm s} \over { M_{\rm p}}} - \alpha_0 \Bigg)^{0.5},
\end{equation}

\noindent where $\alpha_0$ defines the mass ratio below which there is no
 enhancement of nuclear star formation (i.e. gas inflow).  Here, the 
gas accretion efficiency has a maximum of 0.56 for 1:1 halo mergers and 
falls to zero at $\alpha_0$. This parametrization is insensitive to the 
stellar feedback prescription. We use $\alpha$ to define how efficiently 
the merger funnels the galaxy's gas to the
black hole accretion disk. Our equation for $\alpha$ differs only slightly
from the \citet{Cox:2008} work in that we adjust $\alpha_0$ to trigger
gas inflow, while their $\alpha_0=0.09$. 

In particular, we contrasted the black hole growth that would result from
only major mergers with the growth that would occur if minor mergers 
were included as well. Our more conservative criterion, for example, allows 
black holes to accrete gas only if the mass ratio of the host dark 
matter halo is less than 4:1 (this is synonymous with setting 
$\alpha_0=0.25$). This cuts off the gas inflow far earlier than predicted by 
the galaxy merger simulations, but is useful as a conservative estimate and to 
compare to our previous work.

We can set an upper constraint on the final black 
hole mass by allowing gas accretion as long as the merging dark matter 
halos have a mass ratio less than 10:1 (consistent with $\alpha_0=0.1$).
Note that $\alpha_0=0.09$ would imply that mass ratios less than
roughly 11:1 would drive gas inflow~\citep{Cox:2008}, rather than 10:1.
 Adopting $\alpha_0=0.09$ would increase the black hole mass only slightly, 
and since our main goal is to compare to our earlier work, we 
retain $\alpha_0=0.1$.

We used a semi-analytic formalism to calculate the
dynamical friction decay time and subsequent merger timescale
of each black hole once it enters a dark matter halo. 
Now that we have a realistic description of the merger time for 
each black hole within a halo, we can allow the black holes to grow for a 
physically-motivated accretion timescale. The accretion of gas onto both 
incoming and the central black hole starts when the two black holes are 
still widely separated, at the moment of the first 
pericenter passage, and continues until the black holes merge~\citep[c.f.]{Dimatteo:2005, colpi:07}. This sets the 
accretion timescale, t$_{\rm acc}$, as follows: 
t$_{\rm acc}$ = t$_{\rm df}$(r=R$_{\rm vir}$) - t$_{\rm dyn}$(r=R$_{\rm vir}$),
where t$_{\rm df}$(r=R$_{\rm vir}$) is the merger timescale including 
dynamical friction; and t$_{\rm dyn}$(r=R$_{\rm vir}$) is dynamical time 
at virial radius R$_{\rm vir}$, which marks the first pericenter pass 
of the black hole. By stopping the accretion as the
black holes merge, we roughly model the effect of black hole feedback 
in stopping further accretion.

Putting these pieces together, the mass accreted by a black hole during 
t$_{\rm acc}$(r=R$_{\rm vir}$) is:

\begin{equation}
M_{\rm acc}=M_{\rm BH,0} (e^{\frac{\alpha t_{\rm acc}}{t_{\rm sal}}}-1),
\end{equation}

\noindent where M$_{\rm BH,0}$ is initial black hole mass, and $\alpha$ is starburst efficiency~\citep{Cox:2008}, and $t_{\rm sal} \equiv  \epsilon M_{\rm BH,0} c^2 / [(1-\epsilon) L]$, where $\epsilon$  is the radiative efficiency, L is 
the luminosity, and c is the speed of light.
After t$_{\rm df}$, the incoming black hole merges with the SMBH at the 
center and a new SMBH is formed after having accreted gas for
 t$_{\rm acc}$. The accretion time and efficiency
both implicitly encode the large-scale dynamics of the merger and 
the bulk gas accretion into the nuclear region, while $t_{\rm sal}$ 
describes the accretion disk physics. As before, we set t$_{\rm sal}$
to describe sustained Eddington-limited accretion with a efficiency of
$\epsilon = 0.1$~\citep{shakura:73}.

In \citet{KHB:2008bhgrowth}, we found that a gas accretion
triggered by major mergers (4:1) and a Boylan-Kolchin dynamical friction
prescription produced a $4\times 10^6 \, {\rm M}_\odot$ SMBH from a 200 ${\rm M}_\odot$ seed 
in place by z$=5$. We consider this our preferred model as 
it produces a SMBH consistent with the Sgr A$^*$. Note, however,
that the minor merger 
(10:1) Boylan-Kolchin prescription also generates a 
$1.3 \times 10^7 \, {\rm M}_\odot$ black hole from the same seed, 
and can be considered a viable model for an M31-like SMBH.

\section{Calculating the Gravitational Wave Signal}

In order to compare our results with other published results, we
follow the approach of \citet{Sesana:04} to calculate the
gravitational wave signal from the black hole mergers in our
simulation volume. In this section, we provide a brief background on
the theory and outline the method; for a more in-depth description of
gravitational waves from SMBHs see, for example, \cite{Berti:2006}.

The orbital motion of the binary system constitutes will excite a
time-dependent mass quadrupole, which generates gravitational
radiation.  The gravitational wave strain, $h$, measures the strength
of the propagating wave.  The amplitude depends on the comoving
distance to the binary, $d$, the rest frame gravitational wave
frequency, $f_{r}$, and the chirp mass of the binary,
$\mathcal{M}$. The rest-frame gravitational wave frequency is related
to the orbital period, $P$, of the binary, $f_{r}= 2/P$, and the chirp
mass depends on the mass of each binary component as follows:
$\mathcal{M} \equiv (m_1 m_2)^{3/5}/(m_1 + m_2)^{1/5}$. When averaged
over the sky position, polarization, and period, the strain can be
written as:

\begin{equation}
{h} = { {{8 \pi^{2/3}} \over {10^{1/2}}} {{G^{5/3} \mathcal{M}^{5/3}}
    \over {c^4 d}}f_r^{2/3}}.
\end{equation}

This gravitational radiation will cause the binary components to
inspiral on ever tighter and faster orbits. When massive black hole
binaries are widely separated and orbiting at low frequencies, the
gravitational radiation emitted is relatively weak, and therefore the
frequency shift per orbit is minuscule. Most of the evolution is spent
in this phase, where a frequency shift of order unity takes many
orbits to achieve.


As the binary separation slowly shrinks, the gravitational radiation
emitted increases dramatically until the binary coalesces. Close to
coalescence, the binary rapidly sweeps through many frequencies in one
orbit, and at the innermost stable circular orbit (ISCO), the change
in frequency per orbit is of order unity.

We set the minimum observed frequency to $f_{{\rm ISCO}}$ for a test
particle orbiting around a single Schwarzschild black hole. In the
black hole mergers we consider, the mass ratios are always well
outside this test particle limit; we adopt the conventional definition
for $f_{\rm ISCO}$ nonetheless:

 \begin{equation}
   {f_{\rm ISCO}} = {{ {c^3} \over {6^{3/2} \pi G } } { {1} \over
       {(m_1+m_2)} } (1+{\rm z})^{-1}},
\end{equation}
where $c$ and $G$ are the speed of light and gravitational constant,
respectively, $m_{1}$ and $m_{2}$ are the black hole masses, and $z$
is the redshift.

The frequency shift rate in the rest-frame is, to first order:
\begin{equation}
{\dot f_r} = { {{96 \pi^{8/3} G^{5/3}} \over {5 c^{5}}} {{f_r^{11/3}
      \mathcal{M}^{5/3}}}}.
\end{equation}
The orbital periods of million solar mass black hole binaries in the
gravitational radiation regime are on the order of hours, which means
that multi-year observations could accumulate many cycles of an
inspiral at a particular frequency.  LISA, for example, is expected to
observe the sky for at least a year, and projections place the LISA
lifespan at roughly a decade.  Million solar mass black holes orbiting
in the LISA frequency band of $\sim 10^{-4} - 1$ Hz are years or less
from coalescence, so LISA observations will be able to track the
inspiral and coalescence phases of supermassive black hole
binaries. The number of cycles accumulated during the inspiral in an
observation of duration $\tau$ depends on the observed gravitational
wave frequency, $n=f \tau$, where $f=f_r/(1+{\rm z})$.  Note that there are
two regimes: $f > n/ \tau$, where the binary sweeps through many
frequencies in an observation, and $f < n/ \tau$, where the signal
builds from several orbits at one frequency.

The observed characteristic strain, $h_c$, is the strain accumulated 
a single observation:
\begin{eqnarray}
h_c &=& h \sqrt{n} \sim {{1} \over {\sqrt{3} \pi^{2/3}}} {{G^{5/6}
    \mathcal{M}^{5/6}} \over {c^{3/2} d}} f_r^{-1/6} \qquad n<f\tau,
\\ h_c &=& h \sqrt{f \tau} \sim { {{8 \pi^{2/3}} \over {10^{1/2}}}
  {{G^{5/3} \mathcal{M}^{5/3}} \over {c^4 d}}f_r^{7/6}\tau} \qquad
n>f\tau.
\end{eqnarray}
Since we have the component mass and redshift of each black hole
merger in our simulation volume, we can calculate $h$ and $h_c$ over
an observation span $\tau$ before merger. Most black holes in our
volume will merge within a year of reaching the LISA band, so a 3-year
observation window should catch these mergers in the act -- if this
volume is a representative slice of the Universe.  We distinguish
between the mergers directly found in our volume and the mergers
extrapolated throughout the Universe in Table~\ref{tab:first}.

\section{Results}

\subsection{A New Gravitational Wave Source}

\begin{figure}
\vspace{0.5in}
\begin{center}
\includegraphics [width=6.5in,height=4in,angle=0]{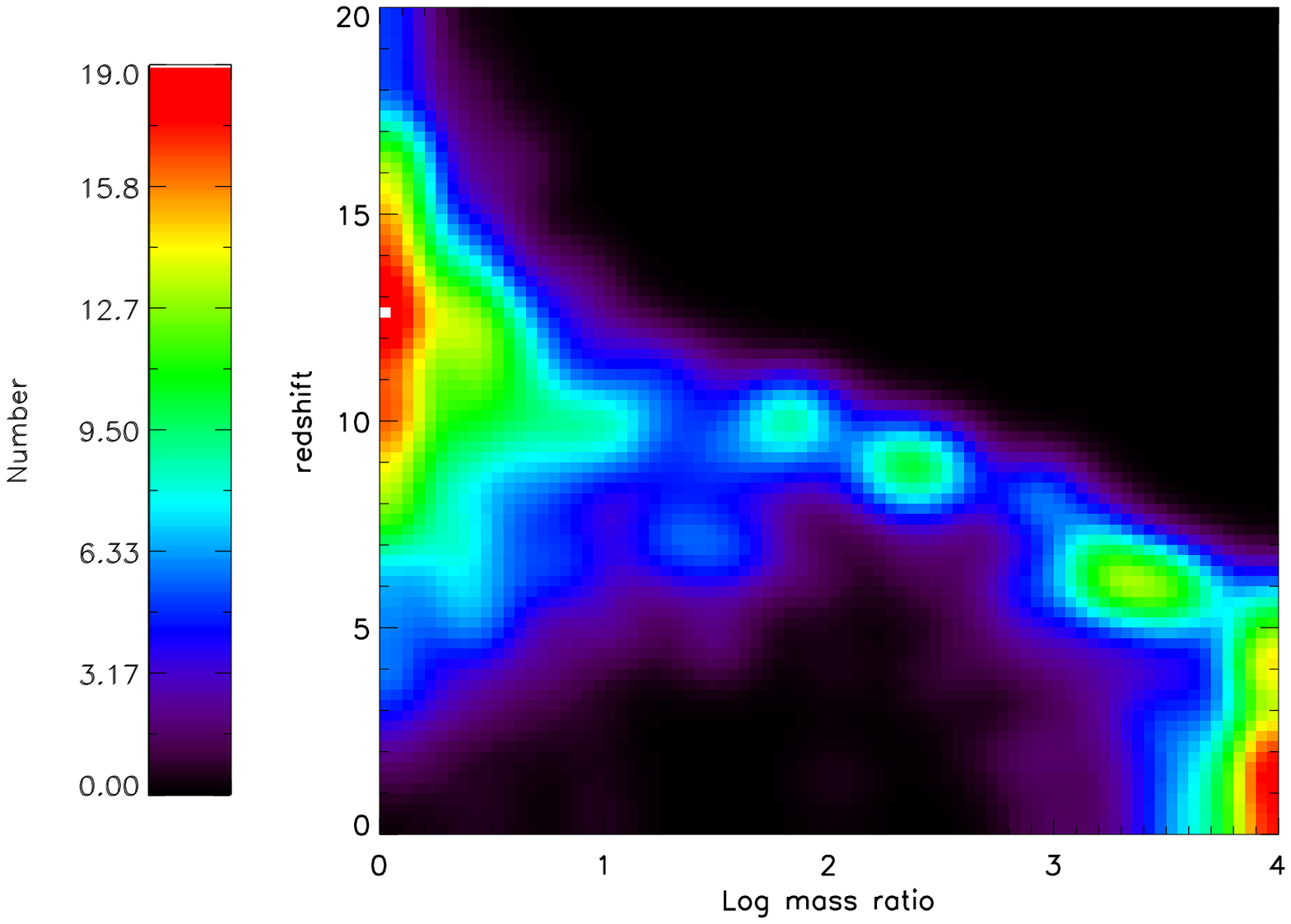}
\caption{Map of the number of black hole mergers in our 1000 Mpc$^3$ volume 
as a function of the log of the black hole mass ratio and the redshift of 
coalescence. The colors represent the number of black hole mergers, with a 
scaling as denoted by the colorbar to the left of the map. Here, the 
dynamical friction is treated with a Boylan-Kolchin formula and gas 
accretion is triggered for dark matter halo msss mergers with mass ratios 
smaller than 4:1 -- major mergers. Note the bimodal distribution, and the fact that most 
mergers are low mass and large mass ratio.}
\label{fig:histmap}
\end{center}
\end{figure}

Overall, most of the mergers in our volume are between seed and
intermediate mass black holes ($O(10^4)\, {\rm M}_\odot$) at redshifts greater
than 5 (Figure~\ref{fig:histmap}). Though these mergers were critical
in assembling the eventual SMBH in our Milky Way analogue, the
generated LISA signal-to-noise ratio of most of these low mass mergers
at high redshift were much less than 1.0. Figure~\ref{fig:histsnr}
demonstrates that the overwhelming majority of black hole mergers will
fall below LISA's detection limit. Although there are a few mergers
that are resolvable by LISA at z$>6$, the number of resolvable sources
increases once the SMBH grows to a mass within the LISA band at
z$<5$, as can be seen in Figure~\ref{fig:histz}. This broadly
agrees with \citet{Sesana:05}, which showed that $\sim10\%$ of the
precursors to $10^9 \, {\rm M}_\odot$ SMBHs are expected to be observable with
LISA out to redshift 10.

\begin{figure}
\vspace{0.5in}
\begin{center}
\includegraphics [width=6.5in,height=4in,angle=0]{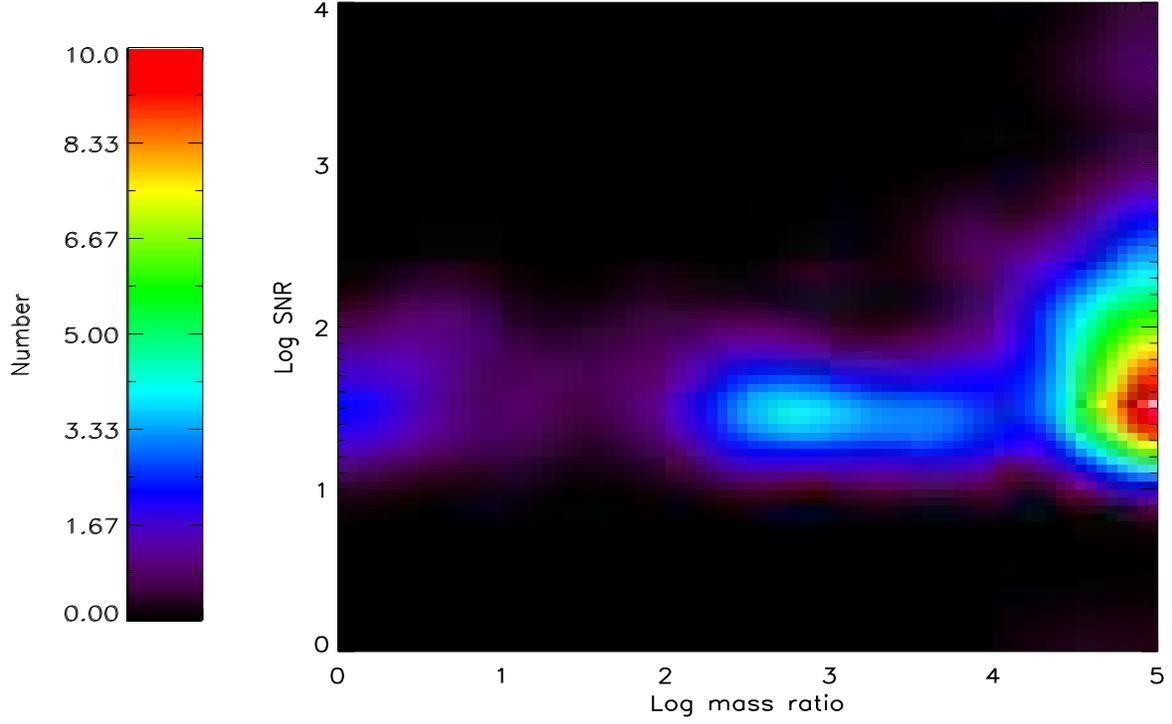}
\caption{Map of the number of {\it resolvable} sources as a function of the log of the black hole mass ratio and the redshift of coalescence. The colorbar to the left translates the color to the number of resolvable sources with a signal-to-noise ratio is greater than 5. Here, the dynamical friction is treated with
a Boylan-Kolchin formula and gas accretion is triggered for dark matter halo
mergers with mass ratios smaller than 10:1 -- minor mergers. Over a 10 year LISA observation, this model is dominated by high mass ratio mergers. }
\label{fig:snrmaprez}
\end{center}
\end{figure}

\begin{figure}
\vspace{0.5in}
\begin{center}
\includegraphics [width=5.in,angle=0]{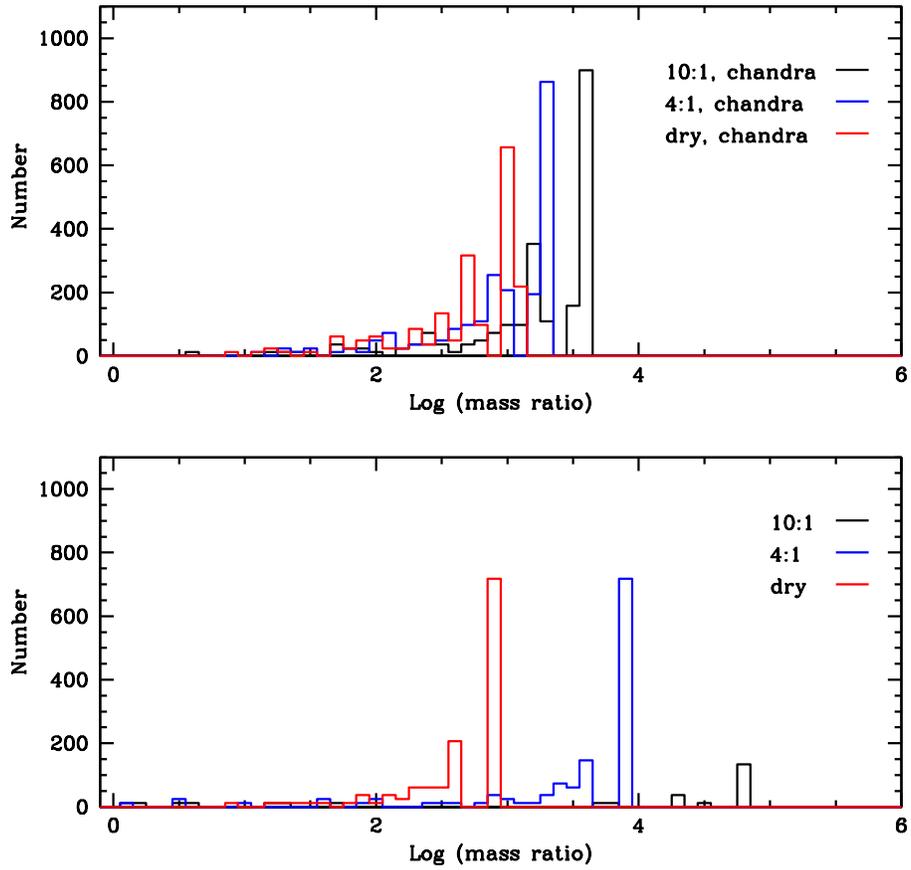}
\caption{Histogram of the mass ratios of sources predicted with a 
signal-to-noise ratio larger than 5, assuming a 3 year LISA observation.
The top panel assumes that the dynamical friction timescale is
set by the Chandrasekhar dynamical friction formula, while the bottom panel
assumes the Boylan-Kolchin formula. The colors correspond to
different black hole growth assumptions: red assumes black holes grow only
through mergers, black assumes that gas accretion onto both black holes is 
triggered if the host dark matter halos have a mass ratio smaller than 10:1, 
and the blue assumes this accretion only occurs for halo mergers with 
4:1 mass ratios. For more detail on the particular black hole growth 
prescription, see Micic, Holley-Bockelmann, and Sigurdsson 2008. 
Regardless of the growth prescription or the large-scale dynamics, we see here that LISA observations will be 
dominated in number by high mass ratio black hole mergers, at least for
the assembly of these relatively light supermassive black holes. }
\label{fig:histq}
\end{center}
\end{figure}

\begin{figure}
\vspace{0.5in}
\begin{center}
\includegraphics [width=5.in,angle=0]{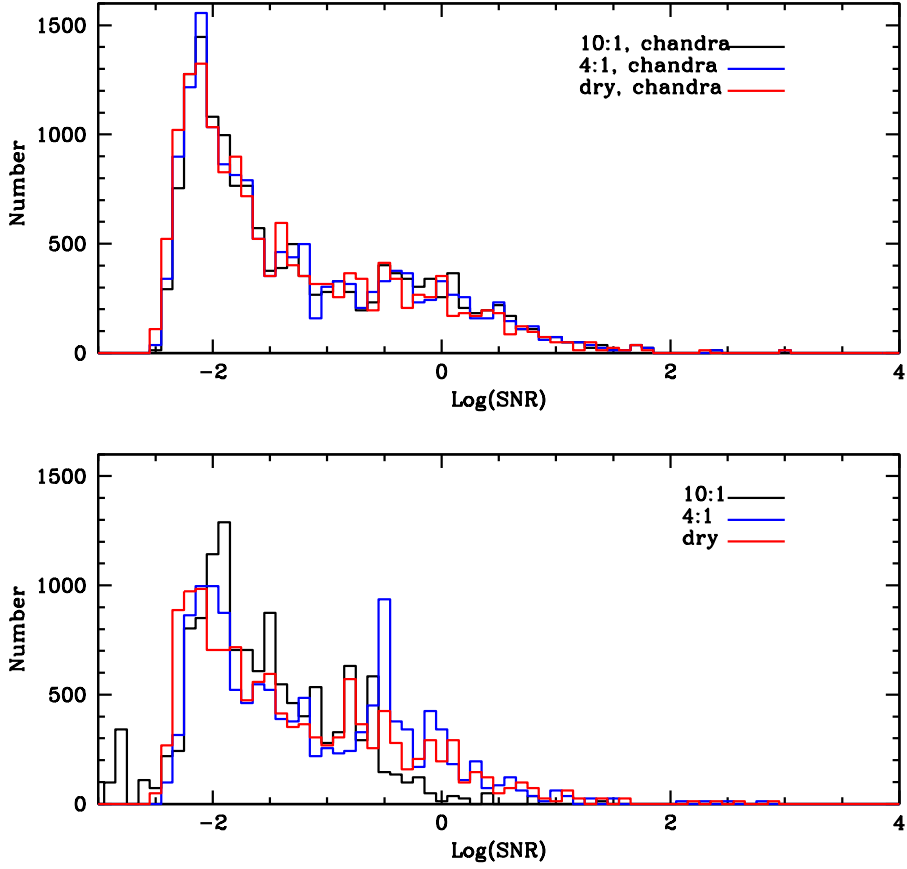}
\caption{Histogram of the signal-to-noise ratio of the 
black hole mergers in our volume (scaled to cosmological volumes) 
for an assumed 3-year LISA observation window. See figure~\ref{fig:histq} for a
description of the panels and colors. Since our cosmological volume is small, 
we are unable to resolve the growth of the most massive black holes. Most of 
our mergers are at high redshift between seed and intermediate mass black holes involved in assembling the light end of the supermassive black hole spectrum;
these high redshift, low mass mergers yield low signal-to-noise sources.}
\label{fig:histsnr}
\end{center}
\end{figure}

\begin{figure}
\vspace{0.5in}
\begin{center}
\includegraphics [width=5.in,angle=0]{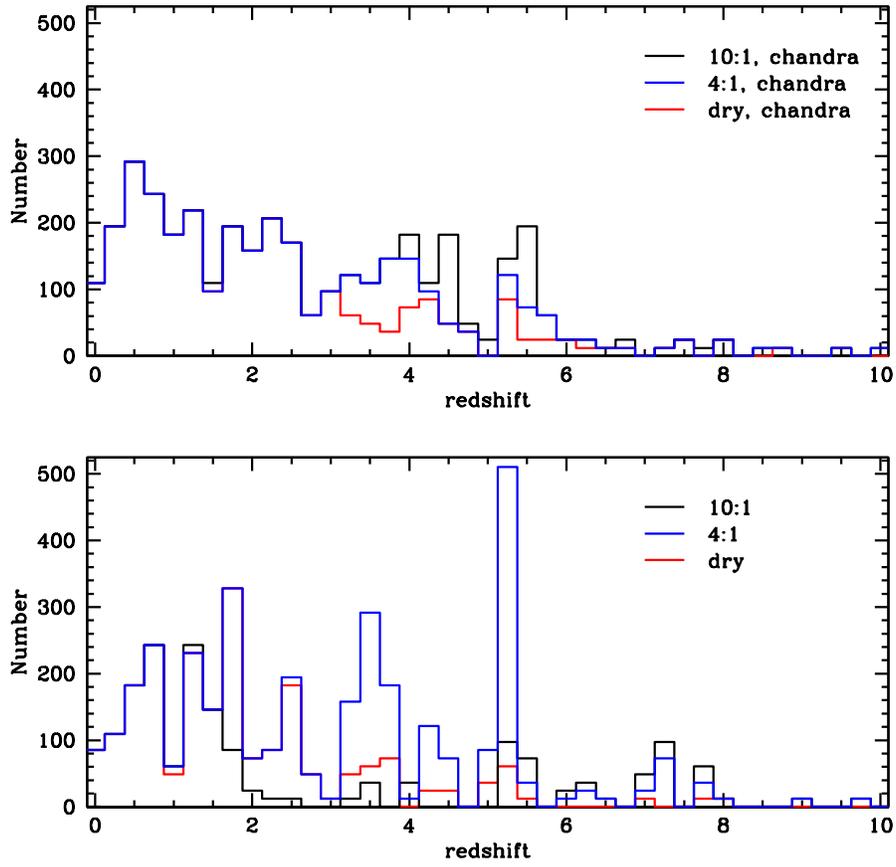}
\caption{Histogram of the redshift of the resolved
black hole mergers in our volume (scaled to cosmological volumes) 
for an assumed 3-year LISA observation window. See figure~\ref{fig:histq} for a
description of the panels and colors. We assume that a signal-to-noise ratio
greater than 5 will be resolvable; given the higher mass ratios
of a typical merger, however, signal-to-noise ratios higher than 30 may be
needed to resolve a source. }
\label{fig:histz}
\end{center}
\end{figure}

However, Figure~\ref{fig:histmap} also reveals a substantial class of
black hole mergers with very high mass ratios; these arise from the
accretion of smaller satellite halos at low redshift.  Since these
mergers are nearby, they are easily detectable with LISA.  This is in
stark contrast to the predictions for assembling the massive end of
the SMBH mass spectrum; \citet{Sesana:07} showed that the detectable
mass ratios are equally distributed in the range of 1-10. In fact,
Figure~\ref{fig:histq} indicates that the most commonly observed black
hole mergers in our volume have mass ratios of 1000 or more (depending
on the black hole growth prescription). Mergers of more equal mass
dark matter halos (and subsequently, the coalescence of more equal
mass black holes) occurred in this volume at $z>8$, and at this epoch,
most of our black holes were too small to be observed by
LISA. Figures~\ref{fig:histmap} and ~\ref{fig:snrmaprez} bear this out
by plotting mapping the number of resolvable sources as a function of
mass ratio and redshift for one black hole growth prescription over a
10 year observation.

\begin{figure}
\vspace{0.5in}
\begin{center}
\includegraphics[width=3.in,angle=0]{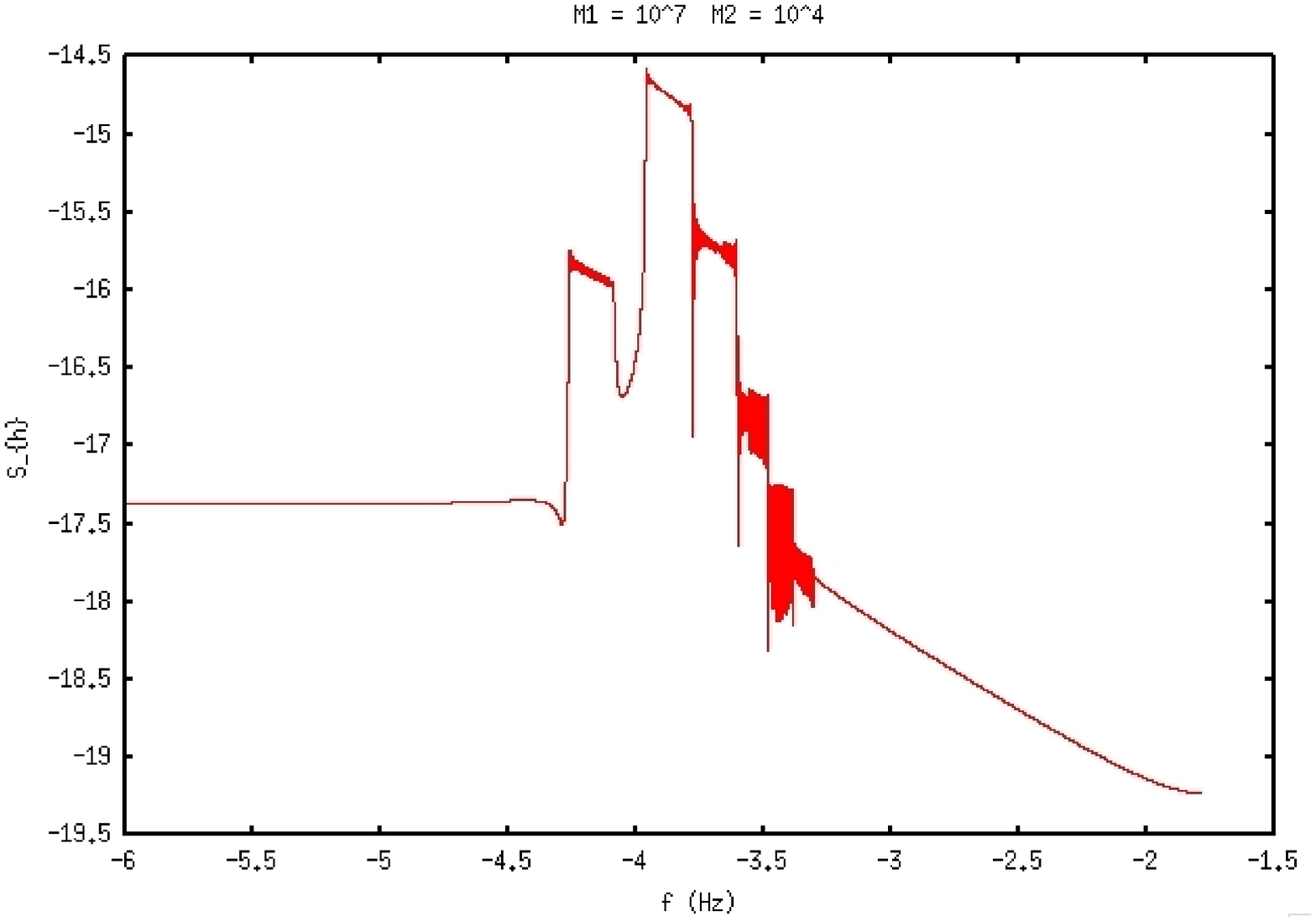}\space\includegraphics[width=3.in,angle=0]{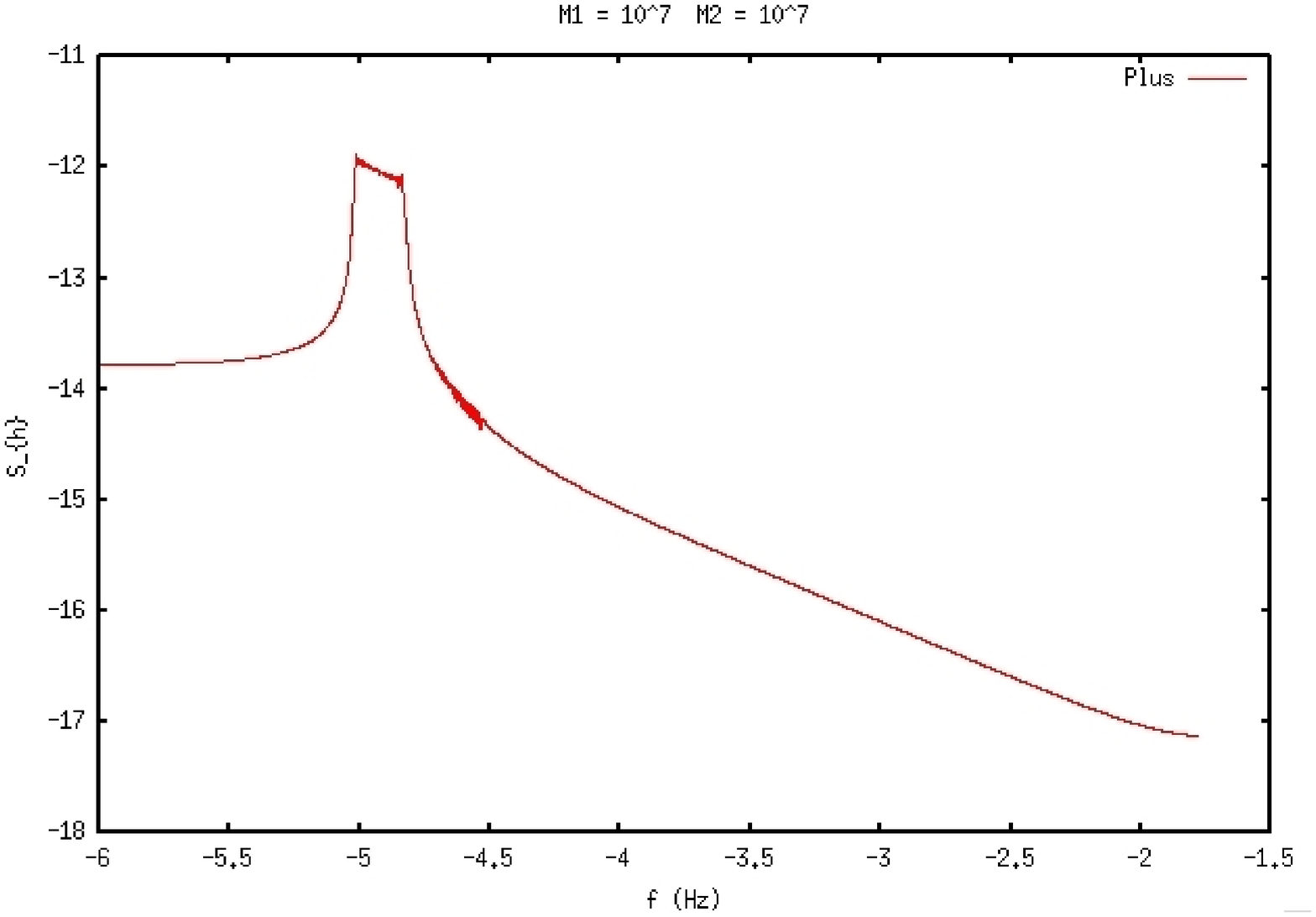}
\caption{ The plus polarization of the gravitational wave amplitude as a function of frequency for two representative black hole mergers at redshift 3. Sources are placed at the same random sky position and orientation. Left: merger with black hole masses $M_1=10^7 M_\odot$ and $M_2=10^4 M_\odot$. Right: merger with equal black hole masses of $10^7 M_\odot$. Notice the broader bandwidth and richer structure of the high mass ratio merger. }
\label{fig:rubbo1}
\end{center}
\end{figure}

The signals from an equal mass inspiraling system and those with mass
ratios of upwards of ten thousand can be quite different.  From an
analytical standpoint, the difference in signals can be understood by
investigating the post-Newtonian expansion of the gravitational wave
strain, which uses the orbital velocity as an expansion parameter.
Each term in the expansion results in harmonics of the orbital
frequency \cite{Blanchet:1996}.  Starting at the first full order,
certain select frequencies are scaled by $\eta \equiv m_{1}m_{2} / (m_{1} +
m_{2})^{2}$, which suppresses those frequencies for systems with large
mass ratios.  Figure~\ref{fig:rubbo1} compares the power spectral
densities for two SMBH systems: one with two $10^7~\textrm{M}_{\odot}$
SMBHs and the other with $m_{1}=10^{7}~\textrm{M}_{\odot}$ and
$m_{2}=10^{4}~\textrm{M}_{\odot}$.  These figures demonstrate how the
equal mass system sweeps through all frequencies while the large mass
ratio system shows much more structure, with many of the frequencies
being suppressed.

The LISA data streams present unique challenges for data mining.
Unlike electromagnetic observatories, LISA is simultaneously senstitive to
gravitational wave sources located all throughout the sky.
LISA will also be sensitive to a wide variety of astrophysical sources -- from
supermassive black hole binaries, to millions of close white 
dwarf binaries within our galaxy, to EMRIs.  The data analysis objective is to
conclusively detect a signal and to extract descriptive parameter
values.  In preparation for the immense data analysis challenge,
simulated data has been produced and distributed to the LISA community
as part of the Mock LISA Data Challenge (MLDC) \cite{Vallisneri:2009}.
So far the challenges have focused on order unity mass ratios for the
simulated supermassive black hole inspirals.  While for most analysis
techniques an assumption about the mass ratio is not hardwired into
the routine, the routines have not been tested at the more extreme
ratios of $10^{5}$ as suggested here.  It is possible that the
surpressed harmonics will cause false-positives or negatives in some
routines because the system will be missed or misidentified.

\subsection{The LISA Signal from Assembling a Milky Way SMBH}

\begin{figure}
\vspace{0.5in}
\begin{center}
\includegraphics [width=5.in,angle=0]{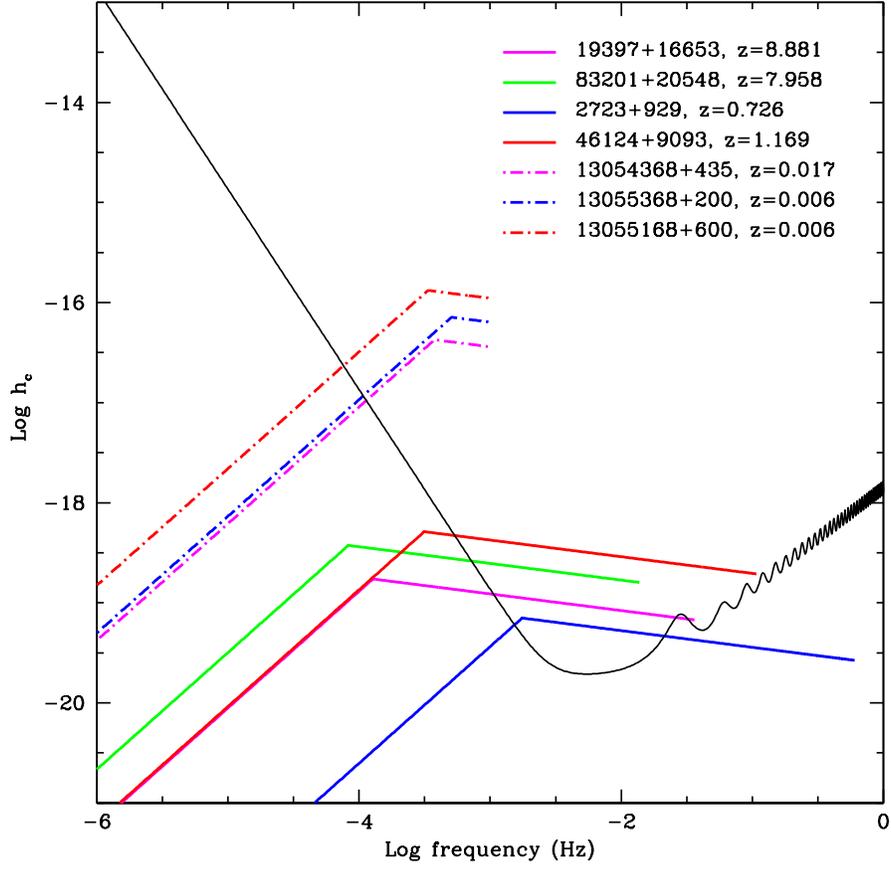}
\caption{Characteristic strain over a 3-year observation
for selected classes of resolvable black hole mergers in the simulation as a function of 
observed frequency. Here, the dynamical friction is treated with
a Boylan-Kolchin formula and gas accretion is triggered for dark matter halo
mergers with mass ratios smaller than 10:1. 
The merger redshift and pre-merger masses of
the black hole binary are labeled; our loudest sources are either caused by
mergers between two intermediate mass black holes, or between a very light seed black hole and the most massive supermassive black hole in our volume.}
\label{fig:tracks}
\end{center}
\end{figure}

Figure~\ref{fig:tracks} presents the total characteristic strain for 
7 bright black hole mergers in our volume, assuming a 
10:1 black hole growth recipe and a Boylan-Kolchin 
dynamical friction treatment. The differences in the two classes of source is
clear -- the brightest sources are high mass ratio mergers which coalesce 
in the LISA band at low redshift, while the other bright class probes
 more equal-mass mergers of low mass black holes at high redshift. Note that
the black holes in the bottom-most track actually merge outside the LISA band
and will be considered an inspiral; this was the only detectable inspiral 
in our volume.

In Table~\ref{tab:snrdist}, we determined the
redshift at which these 7 mergers would become undetectable in 
a 3 year observation.
Understandably, the highly unequal mass mergers can only be detected out to 
a luminosity distance of roughly 3 Gpc, which is similar to the distance 
probed by extreme mass ratio inspirals~\citep{Gair:04}.
 LISA observations of this class of
high mass ratio merger, then, may be useful to probe the fraction of black hole mass is accreted at late times for this low mass SMBH range.

\begin{figure}
\vspace{0.5in}
\begin{center}
\includegraphics [width=5.in,angle=0]{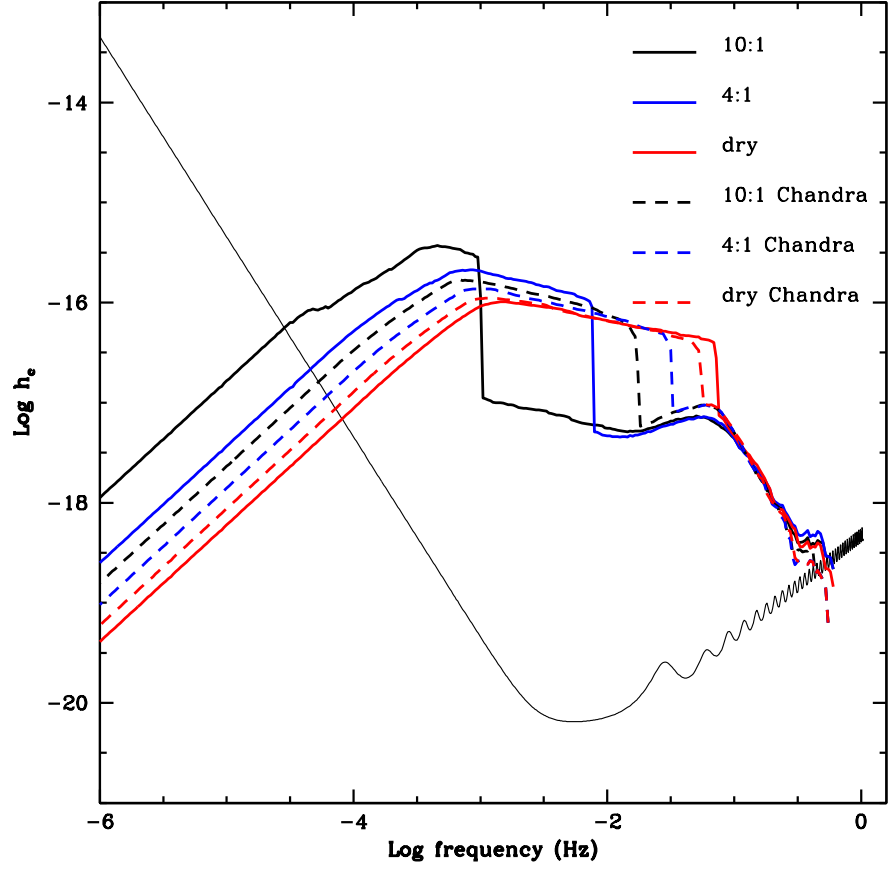}
\caption{ Total characteristic strain as a function of
observed frequency for all the massive black hole mergers in the 
cosmological volume for a three year observation. The solid curve is the
sensitivity curve for LISA for a single-arm Michaelson configuration.}
\label{fig:hsum}
\end{center}
\end{figure}

Figure~\ref{fig:hsum} presents the total characteristic strain for 
the LISA detector
from all the black hole mergers in our 1000 h$^{-3}$ Mpc$^3$ volume over a
3-year observation span. Each curve has a low frequency rise 
that drops steeply off before hitting a
shallow plateau; the rise and drop-off is the signature of the accumulated
high mass ratio mergers, while the shallow segment marks the sum of the 
relatively equal mass mergers. We caution that this implicitly assumes 
that all of these mergers would take place during this 3-year period, or 
alternatively that this volume is representative of the Universe.
Issues of cosmic variance aside, we can see that the black hole 
growth prescription strongly influences the total strain in the 
LISA band; this is best seen when dynamical friction is 
described by~\citet{Boylan:08} where the drop-off changes by a decade in 
frequency in response to a change in the typical mass of the local SMBHs.

\subsection{Extrapolating to Universal Black Hole Merger Rates}

Although this is plagued by cosmic variance, it is instructive to 
estimate the number of mergers observable by LISA in the Universe
expected from assembling these lightest SMBHs. Table~\ref{tab:first} presents the
extrapolated Universal LISA merger rates. One surprising point is that 
the LISA merger rates depend so strongly on the adopted form of
the dynamical friction force. All previous LISA merger rate estimates
have used a semi-analytic technique that employs Chandrasekhar dynamical 
friction to merge the dark matter halos and to usher the
black holes to the inner kiloparsec. However, both
perturbation theory~\citep{colpi:99} and numerical simulations~\citep{weinberg:89,KHB:1999:sat,Boylan:08}
have shown that Chandrasekhar dynamical friction approximates the
merger time to within a factor of two, at best. When we employ
a dynamical friction formalism that is based on fits to merger 
timescales from numerical simulations, we find that the black hole mergers
are delayed to lower redshift. Naively, this would simply make every merger
louder. However, in our SMBH growth prescription, there is an additional 
effect: since the incoming SMBH drives gas inflow to the primary
SMBH over a longer timespan, resulting SMBH ultimately grows more massive
than with Chandrasekhar dynamical friction. This can place the SMBHs in
our volume just out of LISA's 'sweet spot' which will decrease
the signal-to-noise ratio of the more massive black holes. Paradoxically, then,
the LISA event rate for more correct merger times drops by as much as
an order of magnitude. For our most realistic model to assemble the
lightest SMBHs -- gas accretion triggered by major mergers, with a
Boylan-Kolchin dynamical friction merger timescale -- the merger rate drops by
a factor of two. Over a 3 year LISA observation, we should be able to
detect at least 70 SMBH mergers with a signal-to-noise ratio greater
than 30 that are involved in assembling the light end of the SMBH mass 
spectrum in the Universe. If we consider longer LISA observation windows,
the number of observed sources increases, naturally, and for a 10 year 
observation, LISA should detect nearly 500 mergers from the assembly of the lightest SMBHs alone.

\section{Discussion}

By concentrating on a small cosmological volume, 
we have been able to model the gravitational wave sources that 
result from the assembly of a $\sim 10^6 - 10^7 \,{\rm M}_\odot$ black hole in a Milky Way mass halo.
We have calculated the gravitational wave strain for each of the black hole
mergers in our volume, and have determined which of those will be detectable
with LISA. Of the ~ 1500 mergers in our volume, we uncovered 
approximately $300-1200$ mergers detectable with a signal-to-noise 
ratio greater than 5 over a LISA observation span of 3 years.

We found that the most common class of observable black hole
merger in our volume is between a SMBH and IMBH (of 200-2000  M$_\odot$ ) 
at z$<0.05$. These IMBHs originally resided in small dark matter halos 
that merged with the massive primary halo
at high redshift and had very long dynamical friction timescales. Before 
the merger occurs, the incoming IMBH may be observed with the next generation 
of X-ray telescopes as a ULX source with a rate of about $\sim$ 3 
- 7 yr$^{-1}$ for 1 $\leq$ z $\leq$ 5. Because of their potential tie to
observable ULXs, and because this class of source has a different waveform
character than other well known gravitational wave sources, such as 
equal mass mergers, intermediate mass ratio inspirals (IMRIs), 
or extreme mass ratio inspirals (EMRIs), we have nominally dubbed this 
class of source as an Ultra Large Inspirals (ULIs). The other class was IMBH-IMBH
mergers at z$=2-8$.

Note that 
we could not resolve the growth of the most massive dark matter halos with our technique, 
and miss the mergers that arise from 
the assembly of the most massive SMBHs. Therefore, we consider our approach 
to be complementary to studies like~\citet{Sesana:05, Sesana:07} -- we 
believe that the high mass ratio mergers identified here would add to 
the rates found in previous studies that focus on black hole growth in 
present-day halos larger than $\sim 10^{11} \, {\rm M}_\odot$.  However, the 
high mass-ratio channel that we have identified may well dominate the 
LISA black hole merger events. In fact, even our most pessimistic estimate ($\sim 70$ events) yields a comparable number of LISA sources as~\citet{Sesana:05} ($\sim 90$ events).

The signal-to-noise ratio listed in Table~\ref{tab:snrdist} assumes simple 
data analysis techniques. If we were to employ similar sophisticated 
tools as are being developed for the extreme mass 
ratio inspirals, such as matched filtering, it is possible
that this class of source may probe slightly larger distances than EMRIs.
In any event, current data analysis techniques to extract SMBH signals 
from LISA data streams all assume the binaries have mass ratios close to 
unity. As we have shown here, the more complex waveform structure of 
these ULIs may require a different data analysis strategy.

In our volume, 
the SMBH was in place at nearly its final mass at redshift 5, and it was
built by mergers of hundreds of $O(10^2) \, {\rm M}_\odot$ black holes at z$>5$ -- these
mergers were too weak to be detectable by LISA. In fact, the first trace of the
growing SMBH occurs at about redshift 7 for our most aggressive gas accretion
prescription. This may have implications for how well LISA observations can constrain the early growth of these lightest SMBHs.

In this paper, we have neglected gravitational wave recoil, a 
potentially important mechanism that may inhibit black hole growth.
Binary black holes strongly radiate linear
momentum in the form of gravitational waves during the plunge
phase of the inspiral -- resulting in a ``kick'' to the new black hole. 
This, in itself, has long been predicted as a consequence
of an asymmetry in the binary orbit or spin configuration. 
Previous kick velocity estimates, though, were either highly uncertain 
or suggested that the resulting gravitational  wave recoil velocity was 
relatively small, astrophysically speaking. Now, recent 
results indicate the recoil can drive a gravitational wave kick velocity as 
fast as $\sim 4000 \,$ $\kms$~\citep[e.g.][]{Herrmann07a,
Gonzalez:2006md, gonzalez:07b, Baker:2006vn, Koppitz:07kick, Campanelli:2007cg, schnittman:07}. In reality, much smaller values than 
this maximum may be expected in gas-rich galaxies due 
to the alignment of the orbital angular momentum
and spins of both black holes~\citep{tamara:07spin}. 
However, even typical kick velocities ($\sim 200~\kms$) are interestingly 
large when compared to the escape velocity of typical astronomical systems -- 
low mass galaxies, as an example, have an escape velocity of $\sim 200~\kms$~\citep[e.g.][]{KHB:2007recoil}. 
The effect of large kicks combined with low escape velocity from the 
centers of small dark matter halos at high redshift plays a major role in
suppressing the growth of black hole seeds into SMBH.
Even the most massive dark matter halo at 
z$\geq$11 can not retain a black hole that receives 
$\geq$ 150 ${\rm km \,s^{-1}}$ kick~\citep{Merritt:2004xa, micic:2006}. 
We have submitted a companion paper that incorporates the effect of recoil velocity on
the expected merger rates and the growth of a Milky Way-mass SMBH, and find that
if there is no spin alignment mechanism, then a Pop III seed black hole can
reach $10^6 \, {\rm M}_\odot$ only 20$\%$ of the time through merger-driven gas accretion.
We are exploring the effect of recoil on the gravitational wave signal in a forthcoming
paper.

\section*{ACKNOWLEDGMENTS}

\bibliography{references}



\clearpage

\begin{figure}
\vspace{0.5in}
\begin{center}
\includegraphics [width=6.5in,height=4in,angle=0]{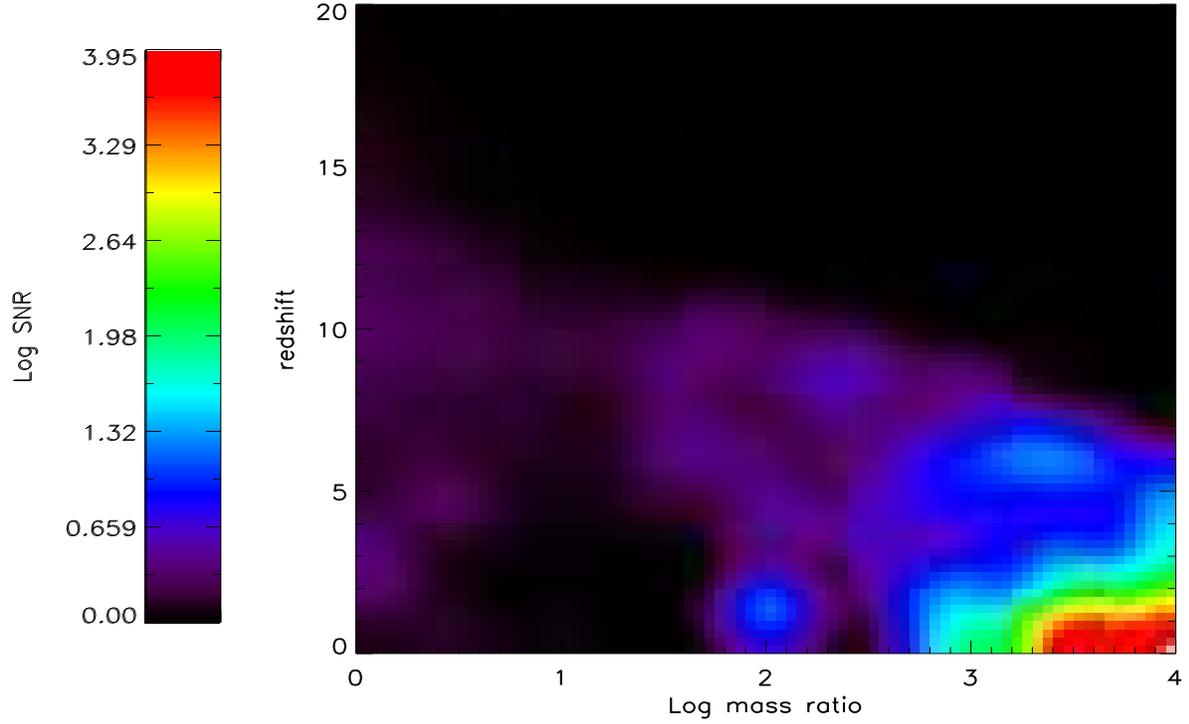}
\caption{Map of the log of the signal-to-noise ratio as a function of the 
log of the black hole mass ratio  and the redshift of coalescence. The 
colorbar to the left translates the colorscale to the log of the SNR -- 
an SNR of 5 would be dark blue. Here, the dynamical friction is treated with
a Boylan-Kolchin formula and gas accretion is triggered for dark matter halo
mergers with mass ratios smaller than 4:1 -- major mergers. Over this 10 year LISA observation,
we can expect to detect no equal mass mergers, but that the loud sources 
will all have mass ratios larger than 100. We reemphasize that this map tracks 
the sources only from the formation channel of the lightest SMBHs.}
\label{fig:snrmapz}
\end{center}
\end{figure}

\clearpage

\begin{table*}
\centering
\caption{Maximum redshift to observe mergers at a signal to noise ratio of 5 in a three year LISA observation.}

\begin{tabular}{@{}cccc@{}}
\hline
\\
$M_{\rm bh1}$
&$M_{\rm bh2}$
& redshift in volume
& max redshift 
\\
\hline\hline
\\
19398 & 16654 & 8.881 & 9.1 \\
83201 & 20548 & 7.958 & 17.5 \\
2723 & 929 & 0.726 & 0.8 \\
46124 & 9093 & 1.169 & 9.8 \\
13054368 & 435 & 0.017 & 0.5\\
13055368 & 200 & 0.006 & 0.4 \\
13055168 & 600 & 0.006 & 0.5 \\

\\
\hline

\label{tab:snrdist}
\end{tabular}

\parbox[s]{4in}{\footnotesize{Column 1: Primary black hole mass. Column 2: Secondary black hole mass. Column 3: Redshift the merger occurred within the simulation volume. Column 4: The redshift to which the merger can be detected at a signal to noise ratio of 5 for a three year LISA observation.}}

\end{table*}

\begin{table*}
\centering
\caption{Number of Resolvable Black Hole Mergers For Several Black Hole Growth 
Scenarios.}
\begin{tabular}{@{}ccccc@{}}
\hline
\\
Gas Accretion Trigger
& Dynamical Friction Type
& SNR
& Observation Time
& Number of Mergers
\\
\hline\hline
\\
Salpeter 10:1 & Chandra & 30 & 10 & 1033\\
Salpeter 4:1 & Chandra & 30 & 10 & 741\\
Dry Mergers & Chandra & 30 & 10 & 608\\
Salpeter 10:1 & Boylan-Kolchin & 30 & 10 & 474\\
Salpeter 4:1 & Boylan-Kolchin & 30 & 10 & 851 \\
Dry Mergers & Boylan-Kolchin & 30 & 10 & 425 \\
\\
Salpeter 10:1 & Chandra & 5 & 10 & 3768\\
Salpeter 4:1 & Chandra & 5 & 10 & 3367\\
Dry Mergers & Chandra & 5 & 10 & 2892\\
Salpeter 10:1 & Boylan-Kolchin & 5 & 10 & 1775\\
Salpeter 4:1 & Boylan-Kolchin & 5 & 10 & 3476\\
Dry Mergers & Boylan-Kolchin & 5 & 10 & 2516\\

\\
\hline

Salpeter 10:1 & Chandra & 30 & 5 & 644\\
Salpeter 4:1 & Chandra & 30 & 5 & 511\\
Dry Mergers & Chandra & 30 & 5 & 400\\
Salpeter 10:1 & Boylan-Kolchin & 30 & 5 & 172\\
Salpeter 4:1 & Boylan-Kolchin & 30 & 5 & 510 \\
Dry Mergers & Boylan-Kolchin & 30 & 5 & 280 \\
\\
Salpeter 10:1 & Chandra & 5 & 5 & 2893\\
Salpeter 4:1 & Chandra & 5 & 5 & 2674\\
Dry Mergers & Chandra & 5 & 5 & 2261\\
Salpeter 10:1 & Boylan-Kolchin & 5 & 5 & 729\\
Salpeter 4:1 & Boylan-Kolchin & 5 & 5 & 2176\\
Dry Mergers & Boylan-Kolchin & 5 & 5 & 1726\\

\\
\hline

Salpeter 10:1 & Chandra & 30 & 3 & 438\\
Salpeter 4:1 & Chandra & 30 & 3 & 401\\
Dry Mergers & Chandra & 30 & 3 & 316\\
Salpeter 10:1 & Boylan-Kolchin & 30 & 3 & 73\\
Salpeter 4:1 & Boylan-Kolchin & 30 & 3 & 316 \\
Dry Mergers & Boylan-Kolchin & 30 & 3 & 194\\
\\
Salpeter 10:1 & Chandra & 5 & 3 & 2200\\
Salpeter 4:1 & Chandra & 5 & 3 & 2323\\
Dry Mergers & Chandra & 5 & 3 & 1898\\
Salpeter 10:1 & Boylan-Kolchin & 5 & 3 & 328\\
Salpeter 4:1 & Boylan-Kolchin & 5 & 3 & 1288\\
Dry Mergers & Boylan-Kolchin & 5 & 3 & 1896\\

\\
\hline

\label{tab:first}
\end{tabular}

\parbox[s]{15.5cm}{
\footnotesize{Columns 1 and 2 describe the black hole growth model; column 1 
indicates which halo mass ratio would trigger gas accretion onto the primary 
black hole; column 2 indicates the dynamical friction formalism that dictates 
the gas accretion timecscale. Column 3: Signal to noise ratio for detection. 
Column 4: Observation duration. Column 5: Predicted number of LISA events.  
Note that column 5 is not the entire LISA event rate; it is the conjectured 
rate from the class of sources involved in assembling the lightest SMBHs. 
Also note that although the predicted event rate varies here by a factor 
of 50, even the most pessimistic rate estimate still will provide ample 
new sources for LISA to observe.}
}

\end{table*}

\end{document}